\documentclass{webofc}
\usepackage[varg]{txfonts}   
\begin{document}
\title{Strange matter in compact stars}
%
%

\author{\firstname{Thomas} \lastname{Kl\"ahn}\inst{1,2}\fnsep\thanks{\email{thomas.klaehn@csulb.edu}} \and
        \firstname{David B.} \lastname{Blaschke}\inst{2,3,4}\fnsep\thanks{\email{david.blaschke@ift.uni.wroc.pl}}\
}

\institute{Department of Physics and Astronomy,
California State University Long Beach,
California 90840, U.S.A
\and
Institute of Theoretical Physics,
University of Wroc\l{}aw,
pl. M. Borna 9, 50-204 Wroclaw, Poland
\and
Bogoliubov Laboratory of Theoretical Physics
JINR Dubna,
Joliot-Curie str. 6, 141980 Dubna, Russia
\and
National Research Nuclear University (MEPhI),
Kashirskoe Shosse 31, 115409 Moscow, Russia
          }

\abstract{%
 We discuss possible scenarios for the existence of strange matter in compact stars.
 The appearance of hyperons leads to a hyperon puzzle in ab-initio approaches based on effective baryon-baryon potentials but is not a severe problem in relativistic mean field models.
In general, the puzzle can be resolved in a natural way if
 hadronic matter gets stiffened at supersaturation densities, an effect based on the quark Pauli quenching between hadrons.
 We explain the conflict between the necessity to implement dynamical chiral symmetry breaking 
 into a model description and the conditions for the appearance of absolutely stable strange quark matter that require both, approximately masslessness of quarks and a mechanism of confinement.
The 
role of strangeness in compact stars (hadronic or quark matter realizations) remains unsettled.
It is not excluded that strangeness plays no role in compact stars at all. 
To answer the question whether the case of absolutely stable strange quark matter can be excluded on theoretical grounds requires an understanding of dense matter that we have not yet reached.
}

\maketitle

\section{Introduction}
\label{sec:intro}

Typically, neutron stars are thought of as exactly this: a gravitationally bound object with a mass of up to two times the mass of our sun that is made of extremely dense nuclear matter,
mostly consisting of neutrons and 
some minor fraction of protons and electrons.
The radius of a neutron star is small, only about 10 to 15 km and consequently the baryon density in a neutron star is huge;
it exceeds the nuclear saturation density $n_S\approx 0.16$ fm$^{-3}$
(where nucleons would start to 'touch'). 
By how much, is not precisely known and certainly depends on the specific object.
However, most calculations agree that in the center of the most massive neutron stars,
densities of four and more times the saturation density can be reached.
This leads to fascinating consequences, among them the possibility 
that nucleons at these extreme densities might deconfine
into a quark--gluon plasma.
Whether this is the case or not has not yet been clarified.
If confirmed, neutron stars would be the only objects in our Universe
where such a transition can take place naturally.
A further interesting consequence of the large densities in a neutron
star is the appearance of particles more massive than neutrons and protons.
In particular, the densities could be large enough to result in particle energies
at the Fermi-surface which exceed the rest mass of hyperons.
Compact stars could contain strange particles,
either in the form of strange hadrons or deconfined strange quarks.
The most exotic scenario is the existence of absolutely stable strange quark matter
\cite{Bodmer:1971we,Witten:1984rs,Farhi:1984qu}.
In the following we briefly sketch these ideas and review current developments
as well as constraints that arise from pulsar measurements.


\section{Neutron Star Masses and Radii}

The more massive a neutron star is, the higher is the central density of this object.
If one therefore assumes interesting things to happen at large densities 
one should
look for the most massive observed neutron stars.
Currently, these are two pulsars with a mass of approximately two solar masses
which have been reported with sufficient precision.
The probably cleanest result has been obtained from the 
Shapiro delay and orbital data of PSR J1614-2230 \cite{Demorest:2010bx}, 
a neutron star in a binary system with a white dwarf which results in a neutron star mass 
of $1.928\pm 0.017$  M$_\odot$ \cite{Fonseca:2016tux}.
An even larger, precise neutron star mass of $2.01\pm0.04$ M$_\odot$ 
has been obtained for PSR J0348 + 0432  \cite{Antoniadis:2013pzd}, another neutron star in 
a binary system with a white dwarf, based on the analysis of orbital data and modeling
the structure of the white dwarf.
More data indicating massive (even heavier) neutron stars are available, 
however not with comparable precision.
It should be noted that the radius and mass of these heavy neutron stars
is almost independent of the symmetry energy and hence the fraction
of particles with different isospin as neutrons and protons.

Considering the distance together with the small radii of neutron stars somewhere between
8 and 16 km, it seems evident that precise radius measurements are challenging.
This is unfortunate, as the radius of a typical neutron with $1.3-1.5$ M$_\odot$ depends
significantly on the symmetry energy. Consequently, precise measurements would be helpful
to constrain the density behavior of this quantity.
Part of the problem is that radius measurements depend on light curve analyses
in various forms and hence are typically highly model dependent.
The ongoing NICER experiment \cite{NICER1} and future missions as ATHENA+ or LOFT 
have been designed to gain more precise and conclusive results.
Together with the recent first observation of gravitational waves in the binary neutron star merger 
GW170817 and the electromagnetic signals from the related kilonova event 
\cite{TheLIGOScientific:2017qsa}
there is hope that our knowledge concerning neutron star masses and radii improves 
significantly within the next few years.

\section{Strange Matter}
\subsection{Hyperons}

At densities of two to three times saturation density the hyperon threshold is expected to be
crossed. 
It has been shown within Brueckner-Bethe-Goldstone calculations \cite{Baldo:1999rq}
that under conservative assumptions 
for the forces involving hyperons the appearance of these additional degrees of freedom softens the equation of state and hence lowers the maximum mass of a neutron star so that even the well-constrained binary radio pulsar masses of typically $\sim1.4~M_\odot$ cannot be reached.
Without additional repulsive hyperon interactions this is exactly what all neutron star calculations based 
on nuclear hyperon model equations of state predict.
This is called the "hyperon puzzle" \cite{Zdunik:2012dj}.
In general, the equation of state beyond saturation density is not well constrained~\cite{Oertel:2016bki}.
For hyperons and their interaction with nucleons and themselves the situation is even worse.
However, from the observed existence of hyperons 
and
massive neutron stars 
it seems evident that nucleon-hyperon and likely hyperon-hyperon repulsion \cite{Rijken:2016uon}
plays an important role
(unless hyperons themselves play no decisive role for neutron star structure).
Meanwhile, different approaches have been successfully applied which 
allow for stable neutron stars with up to two solar masses.
Repulsion stiffens the hyperon matter equation of state
sufficiently to account for massive neutron stars. At the same time the stiffening results in
an onset of hyperon degrees of freedom at higher densities.
Thus, the fraction of hyperons is reduced.
In another scenario this idea is taken to the limit where the hyperon onset density is
larger than the densities one would find in a neutron star.
Generally, in relativistic mean-field models for hypernuclear matter  \cite{Hofmann:2000mc}
there is no severe hyperon puzzle,
in particular due to the sufficiently repulsive effect of the $\phi-$meson mean field
\cite{Weissenborn:2011ut}. 
Thos holds also for hypermatter stars with (color supercondicting) quark matter cores
\cite{Bonanno:2011ch,Lastowiecki:2011hh}.
A third solution to the hyperon problem is a transition to sufficiently stiff quark matter \cite{Baldo:2003vx}
which can happen before or after the transition to hyperon matter.

\subsection{Quark Matter}
Before discussing strange quark matter degrees of freedom we briefly describe
the challenges in describing deconfined quark matter.
Quarks, gluons and their interactions are synonymous for QCD, the gauge field theory of the strong interaction. 
At high densities and zero temperature (as in a neutron star) lattice QCD, the ab initio
approach to QCD still fails.
Although one can hope that perturbative results will constrain effective models at large
densities 
the perturbative domain does not overlap with the densities one expects in a neutron star 
\cite{Kurkela:2014vha}. 
Moreover, the QCD phase transition from confined hadrons to deconfined quarks
is characterized by features which are not accessible by perturbative approaches to QCD.
A standard example for this statement is the description of chiral symmetry breaking
which in the light quark sector is a distinctively non-perturbative feature.
The most common approach to avoid these difficulties is to apply effective quark matter models 
which are not derived from QCD but 
account for
certain characteristic features, most notably again the dynamical breaking of chiral symmetry.
Prominent examples for effective models are the thermodynamic bag model as the high density limit of the MIT bag model,
and effective relativistic mean field models, typically of the Nambu--Jona-Lasinio type
\cite{Buballa:2003qv}.
A next approach which is well developed for the study of hadrons is the non-perturbative Dyson-Schwinger formalism
which starts from the QCD action and derives gap equations to determine QCD's n-point Green-functions in a medium.
As these typically couple to higher order Green-functions, truncation schemes are introduced
which, if chosen wisely, preserve key features of QCD \cite{Cloet:2013jya}.

Quark matter at finite densities and zero or small temperature can exhibit an extremely rich
phase structure due to different pairing mechanisms which arise from the coupling of
color, flavor and spin degrees of freedom and result in a variety of different possible condensates.
The importance of condensates is illustrated by the color-flavor locked phase which can appear 
in three flavor (up, down, strange) matter and is shown to be the asymptotic ground state of quark matter 
at low temperature \cite{Schafer:1999jg}.
As described for hyperon matter, quark matter would be too soft to account for massive neutron
stars if repulsive interactions would not be taken into account. 
However, vector repulsion arises as naturally
as the breaking of chiral symmetry in effective relativistic models and in the Dyson-Schwinger approach
as dressing of the fermion propagator's vector and scalar part, respectively.
The thermodynamic bag model does not account for vector interactions.
Historically, one can argue that this is the heritage of the MIT bag model which has been developed to
describe hadron properties in vacuum where terms that scale with the density are not relevant.
However, taking perturbative corrections to the thermodynamic bag model into account,
it can be made sufficiently stiff to account for massive neutron stars. 
It should be noted though, that the expansion parameter $\alpha_S$ cannot be considered small in this case.

Chiral symmetry breaking, the generation of an effective quark mass due to quark gluon interaction
is a crucial effect as will be shown in more detail later.
For now, it is worth to highlight that the restoration of chiral symmetry is flavor dependent. 
This can result in a sequential appearance of quark flavors at different densities 
\cite{Blaschke:2008br,Alford:2017qgh}. 
In particular if the three quark flavors are uncoupled, strange quarks,
due to higher dressed masses, will appear at higher densities than light quarks.
Even scenarios with a sequential appearance of first up, then down, then strange quarks have been discussed.
With the appearance of condensates such a general statement is not feasible.
Different studies suggest different scenarios. 
The transition densities for all flavors can coincide or differ widely.
Examples of the first case are the thermodynamic bag model (because it neglects chiral symmetry breaking and thus reduces the onset density of all quark flavors) and a Dyson-Schwinger study 
\cite{Nickel:2006vf} where flavor coupling results in a simultaneous appearance of all flavors.
NJL type models in general predict a sequential transition.
However, in all these scenarios the light quark threshold is a lower bound for the strange quark threshold.

\subsection{Strange Quark Matter in Neutron Star Cores}
We first discuss the situation where normal nuclear or hyperon matter deconfines at a critical density and forms a quark-gluon plasma.
This problem is still far from being solved consistently in a way where hadrons would be taken into account as actually confined quarks which then deconfine dynamically.
The standard way to circumvent a detailed description of deconfinement is to choose a two phase approach
where nuclear and quark matter are modeled independently and the transition is constructed thermodynamically consistently, viz. in terms of a Maxwell (or similar) construction.
In case of the Maxwell construction one simply determines at which baryon-chemical potential 
the pressure of the nuclear and quark phase are equal and thus defines the transition point.
By construction, this implies a first order transition which 'switches' from a given nuclear equation of state to
a softer quark matter equation of state. Therefore, this procedure requires a nuclear EoS which is stiff enough
to support at least a two solar mass neutron star, and a quark matter EoS which is softer but stiff enough to do the same.
How exactly this happens can vary. 
The quark matter EoS can mimic the nuclear EoS, be generally softer or at some
density can turn even stiffer than the underlying nuclear EoS
so that a second crossing of the pressure curves occurs ("reconfinement problem" \cite{Zdunik:2012dj}). 
The latter scenario is justified if one assumes that
at densities far enough beyond the transition a comparison of both phases is meaningless as the nuclear EoS does not describe any physical reality anymore.
In any of these scenarios, repulsion is a crucial feature to account for the existence of massive neutron stars and,
as stated earlier, a natural property of relativistic models.
For the onset density of hyperon matter we discussed how it is pushed to increasingly high densities with increasing 
stiffness or repulsion.
The same would be true for quark matter if condensates are not taken into account.
However, condensates couple colors and flavors and can lower
the transition density. 
Therefore, quark models can account for transition densities far below the hyperon threshold.

For the appearance of strange matter in neutron stars, a couple of scenarios emerge which we divide into two groups.
First, nuclear matter can deconfine directly into (three flavor) strange matter as one would find it for the
thermodynamical bag model, as described earlier.
Second, a sequential transition from nuclear to two flavor followed by a transition to three flavor matter takes place.
Similar to this scenario, there are two more cases which would not result in a strange matter core but are not less realistic.
The sequential transition results in neutron star configurations which are stable at all densities
below the strange quark threshold but unstable beyond due to the softening of the EoS with this new degree of freedom.
This situation has been described for an NJL model with diquark couplings 
\cite{Klahn:2006iw}.
However, choosing a finite constant as offset to the quark pressure
can alter this result \cite{Bonanno:2011ch} and render neutron stars with strange matter core stable 
(the thermodynamic properties of matter are described in terms of pressure derivatives, hence a constant offset keeps them intact).
Another way to stabilize strange core configurations is a strongly density dependent stiffening
of the strange matter EoS following the transition
\cite{Benic:2014jia,Kaltenborn:2017hus}.
This can result in situations where stable two flavor and three flavor quark core neutron star configurations
are separated by a population gap at intermediate central neutron star densities.
In extreme scenarios, this can generate separated mass twin configurations, viz.
two neutron star families with similar masses but very different radii 
\cite{Blaschke:2013ana,Alvarez-Castillo:2017qki}.
If members of both families could be observed, this would be a strong indicator
for a first order QCD-like phase transition where (the smaller) neutron stars can carry a core made of strange matter \cite{Bastian:2017fzo}.

Another indicator for a first order phase transition in dense matter would be the observation of a delayed second neutrino signal after a supernova. 
This has been suggested based on simulations which applied a bag model \cite{Sagert:2008ka}.
Although such a measurement would be very exciting it is not clear how it would address the question whether the transition involved strange matter, or quark matter at all.

\subsection{Absolutely Stable Strange Matter?}
The hypothesis that strange matter could be absolutely stable, bases 
on the observation that the appearance of strange quarks lowers the energy per baryon.
As two flavor quark matter is evidently less stable than Fe (otherwise Fe would decay into it's quark components and so would we)
this leaves a window where two-flavor matter is less and strange matter more stable than iron.
This hypothesis has been supported by certain parametrizations of the thermodynamic bag model.
Choosing the proper bag constant one can indeed find exactly the proposed situation.
If this scenario is reality it would have a number of interesting consequences.
Therefore, the search for stable strange matter inspired a multitude of experiments and has born many new ideas.
Strange matter could form strange nuggets of extreme density with rather small atomic numbers and hence
extremely low cross sections. It could form objects very similar to a neutron star, almost entirely made of strange matter, see~\cite{Bombaci:2001uk,Weber:2004kj} for reviews of this and other scenarios involving strangeness in compact stars.
A seed of strange matter in a neutron star could destabilize the surrounding matter and thus trigger a conversion of the neutron star interior into strange matter. 
Recently, it has been proposed that muonic bundles that have been observed at ALICE (CERN) are produced by strangelets \cite{Kankiewicz:2017wbo}.
Currently, other ways to detect strangelets are actively investigated \cite{VanDevender:2017avk}.
The idea of absolutely stable strange matter is certainly appealing.

\begin{figure}
\centering
\sidecaption
\includegraphics[width=8cm,clip]{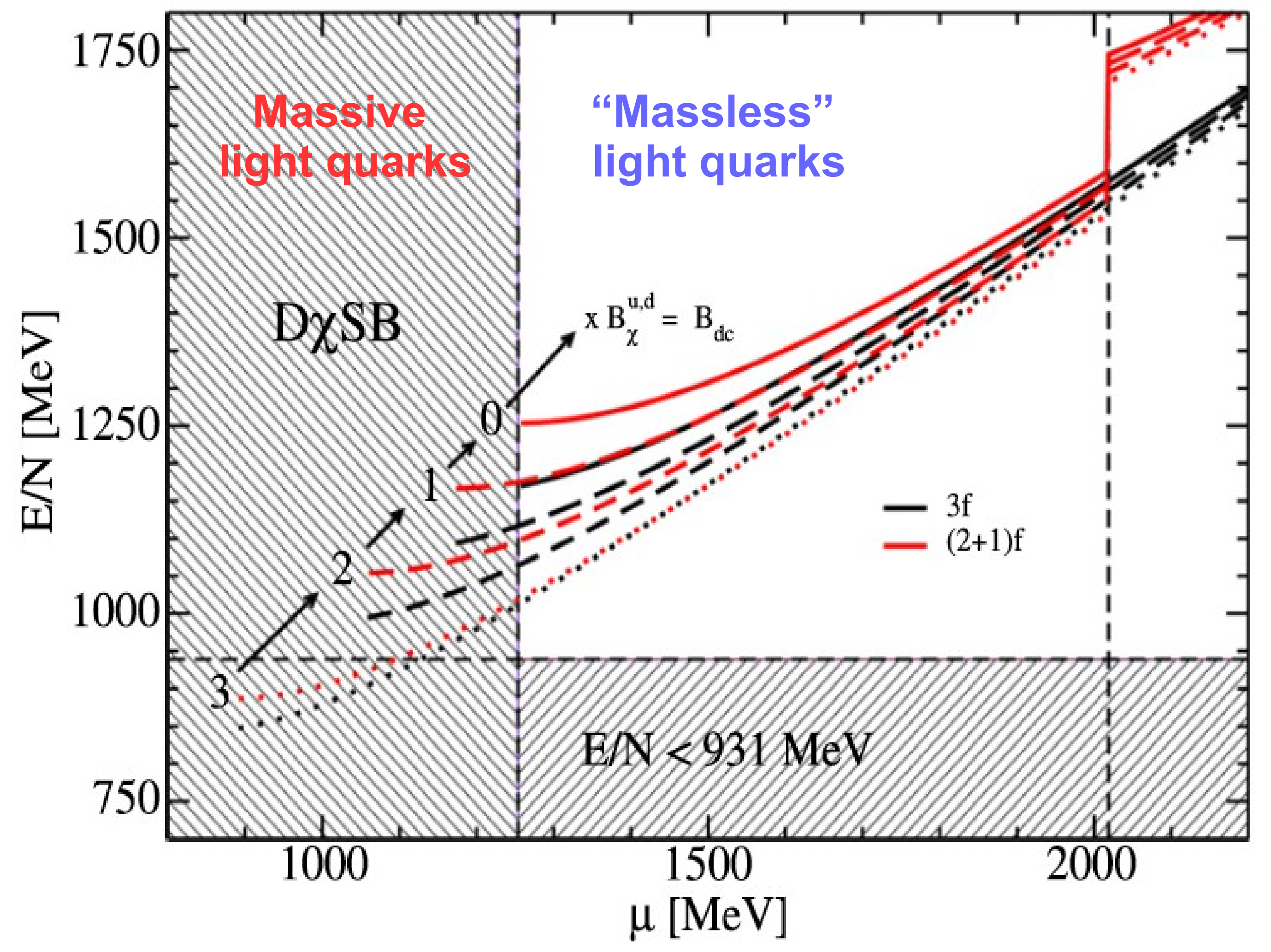}
\caption{Energy per baryon vs. baryon chemical potential for the vBag quark matter model
\cite{Klahn:2015mfa}. 
Absolutely stable strange quark matter could be obtained if dynamical chiral symmetry breaking (D$\chi$SB) is neglected. However: The vertical band  marks the approximate region where chiral symmetry is broken. For the shown curves D$\chi$SB is ignored for all (black) or all but the s-quark (red).
At densities below the s-quark threshold this allows to effectively compare two(red)- and three-flavor(black) quark matter for different effective
bag constants. In none of the cases strange matter is more stable than iron (E/N<931 MeV) in a density domain where chiral symmetry is restored.}
\label{fig-1}       
\end{figure}

Theoretically, as mentioned before, the hypothesis has been based on the thermodynamic bag model.
It should be noticed, that this model lacks a key feature of QCD, namely chiral symmetry breaking, 
viz. quarks in the thermodynamic bag model are assumed to have bare masses and therefore
extremely small threshold densities (the critical chemical potential scales with the effective mass).
NJL-type models, which do generate dressed quark masses, do not confirm that strange
matter is absolutely stable. 
The reason for this is easily found: In the (low) density domain where the bag model predicts
absolutely stable strange matter, NJL type models find chiral symmetry to be broken, hence significantly
larger quark masses and consequently a higher energy per particle, too high to render strange matter absolutely stable, see Fig.~\ref{fig-1}. 
An appealing and sometimes confusing feature of the thermodynamic bag model is that it originates from
the MIT bag model which has been developed to describe hadron properties. 
This evidently seems possible even though the model assumes bare quark masses
and makes it easier to believe that massless quarks could form stable strangelets.
This confusion can be cleared if one realizes that the MIT bag model does not only
assume bare quark masses but a bag constant which is introduced as synonym for
'all we don't know about confinement and further interactions'. 
The fathers of the MIT bag model stated this explicitly.
In \cite{Klahn:2015mfa} we illustrated how to translate a model with chiral symmetry breaking
into a bag type model and where this model would break, which is when chiral symmetry is broken.
For an earlier investigation of this kind, see \cite{Buballa:2003qv} for the NJL model case and  
\cite{Gocke:2001ri} for a nonlocal chiral quark model. 
Effects of color superconductivity in this context have been studied and parametrized in 
\cite{Grigorian:2003vi}.
%

%
The bag constant mostly originates from chiral symmetry breaking. It is subtracted because the difference between the vacuum pressure of massive quarks and effectively massless quarks is negative. The absolute value is reduced by confinement. Thinking of the absolute value of the bag constant as the energy of a hadron, this makes sense as 
confinement reduces the energy of the system of chirally broken quarks by the binding energy.
The MIT bag model is not consistent in the sense that it ignores any relation
between quark mass and bag constant. In vacuum this is a reasonable approximation for two reasons:
First, the model describes hadrons and does not attempt to predict any dynamical property of individual constituent quarks.
Second, it can be fitted to observables and thus repairs the inherent shortcomings.
The thermodynamic bag model addresses both effects, chiral symmetry breaking and confinement, in terms of one parameter - the bag constant. Statements regarding quark matter based on the thermodynamic bag model, in particular at low densities where the bag constant affects the total pressure significantly, should be considered with considerable caution as it is all but clear, that the model actually describes deconfined matter.

It should be noticed, that effective quark models with density dependent quark masses 
have been suggested which indeed would predict absolutely stable strange matter
\cite{Dondi:2016yjl}.
A distinct feature of these models is a steep {\it concave} decrease of the strange quark mass at
comparably low density opposing to the typically {\it convex} behavior. 
This reduces the effective quark mass drastically already at low densities 
which makes it very similar to to the thermodynamic bag model,
 evidently with similar results regarding the stability of strange matter.
It would be interesting to see, how a microscopic approach which generates
this kind of density behavior would perform describing hadron properties.

\section{Conclusions}
We discussed the existence of strange matter in compact stars in hadronic and quark matter phases.
The appearance of hyperons leads to a hyperon puzzle in approaches based on effective 
baryon-baryon potentials but is not a severe problem in relativistic mean field models.
The puzzle is resolved for a stiffening of hadronic matter at supersaturation densities, 
an effect based on the quark Pauli quenching between hadrons.
We further outlined the conflict between the necessity to implement dynamical chiral symmetry breaking 
for a realistic quark matter model and the condition of undressed, approximate massless quarks for the appearance of absolutely stable strange quark matter. 
The existence of absolutely stable strange quark matter cannot be excluded on theoretical grounds only.
However, we outlined the problems of the reasoning that lead to this hypothesis.
In general, the role of strangeness in compact stars in hadronic or quark matter realizations remains unsettled.
The possibilities range from the existence of pure strange stars to 
scenarios where strange particles play no role in compact stars at all.

\section{Acknowledgement}
This work was supported by the Polish National Science Center (NCN) under grant No. 
UMO-2014/13/B/ST9/02621 (T.K.) and 
by the Russian Science Foundation under grant No.  17-12-01427 (D.B.).


\begin{thebibliography}{}
%
%

\bibitem{Bodmer:1971we}
  A.~R.~Bodmer,
  Phys.\ Rev.\ D {\bf 4} (1971) 1601.

\bibitem{Witten:1984rs}
  E.~Witten,
  Phys.\ Rev.\ D {\bf 30} (1984) 272.

\bibitem{Farhi:1984qu}
  E.~Farhi and R.~L.~Jaffe,
  Phys.\ Rev.\ D {\bf 30} (1984) 2379.

\bibitem{Demorest:2010bx}
P.~Demorest, T.~Pennucci, S.~Ransom, M.~Roberts and J.~Hessels,
Nature {\bf 467} (2010) 1081.

\bibitem{Fonseca:2016tux} 
  E.~Fonseca {\it et al.},
  Astrophys.\ J.\  {\bf 832} (2016) 167.
  
\bibitem{Antoniadis:2013pzd}
  J.~Antoniadis {\it et al.},
  Science {\bf 340} (2013) 6131.

\bibitem{NICER1}
\newblock https://heasarc.gsfc.nasa.gov/docs/nicer

\bibitem{TheLIGOScientific:2017qsa}
  B.~P.~Abbott {\it et al.} [LIGO Scientific and Virgo Collaborations],
  Phys.\ Rev.\ Lett.\  {\bf 119} (2017) no.16,  161101.

\bibitem{Baldo:1999rq}
  M.~Baldo, G.~F.~Burgio and H.~J.~Schulze,
  Phys.\ Rev.\ C {\bf 61} (2000) 055801.
  
\bibitem{Zdunik:2012dj}
  J.~L.~Zdunik and P.~Haensel,
  Astron.\ Astrophys.\  {\bf 551} (2013) A61.

\bibitem{Oertel:2016bki} 
  M.~Oertel, M.~Hempel, T.~Kl\"ahn and S.~Typel,
  Rev.\ Mod.\ Phys.\  {\bf 89}, no. 1, 015007 (2017)

\bibitem{Bombaci:2001uk} 
  I.~Bombaci,
  Lect.\ Notes Phys.\  {\bf 578} (2001) 253.

\bibitem{Weber:2004kj}
  F.~Weber,
  Prog.\ Part.\ Nucl.\ Phys.\  {\bf 54} (2005) 193


\bibitem{Rijken:2016uon}
  T.~A.~Rijken and H.~J.~Schulze,
  Eur.\ Phys.\ J.\ A {\bf 52} (2016) no.2,  21.

\bibitem{Hofmann:2000mc}
  F.~Hofmann, C.~M.~Keil and H.~Lenske,
  Phys.\ Rev.\ C {\bf 64} (2001) 025804.
  
\bibitem{Weissenborn:2011ut}
  S.~Weissenborn, D.~Chatterjee and J.~Schaffner-Bielich,
  Phys.\ Rev.\ C {\bf 85} (2012) no.6,  065802
   Erratum: [Phys.\ Rev.\ C {\bf 90} (2014) no.1,  019904].
  
\bibitem{Bonanno:2011ch}
  L.~Bonanno and A.~Sedrakian,
  Astron.\ Astrophys.\  {\bf 539} (2012) A16.
  
\bibitem{Lastowiecki:2011hh}
  R.~Lastowiecki, D.~Blaschke, H.~Grigorian and S.~Typel,
  Acta Phys.\ Polon.\ Supp.\  {\bf 5} (2012) 535.

\bibitem{Baldo:2003vx}
  M.~Baldo, G.~F.~Burgio and H.-J.~Schulze,
  astro-ph/0312446.
  
\bibitem{Kurkela:2014vha}
  A.~Kurkela, E.~S.~Fraga, J.~Schaffner-Bielich and A.~Vuorinen,
  Astrophys.\ J.\  {\bf 789} (2014) 127.

\bibitem{Buballa:2003qv}
  M.~Buballa,
  Phys.\ Rept.\  {\bf 407} (2005) 205.

\bibitem{Cloet:2013jya}
  I.~C.~Cloet and C.~D.~Roberts,
  Prog.\ Part.\ Nucl.\ Phys.\  {\bf 77} (2014) 1.

\bibitem{Schafer:1999jg}
  T.~Sch\"afer and F.~Wilczek,
  Phys.\ Rev.\ D {\bf 60} (1999) 114033.

\bibitem{Blaschke:2008br}
  D.~Blaschke, F.~Sandin, T.~Klahn and J.~Berdermann,
  Phys.\ Rev.\ C {\bf 80} (2009) 065807.

\bibitem{Alford:2017qgh}
  M.~G.~Alford and A.~Sedrakian,
  Phys.\ Rev.\ Lett.\  {\bf 119} (2017) no.16,  161104.
  
\bibitem{Nickel:2006vf}
  D.~Nickel, J.~Wambach and R.~Alkofer,
  Phys.\ Rev.\ D {\bf 73} (2006) 114028.

\bibitem{Klahn:2006iw}
  T.~Kl\"ahn, D.~Blaschke, F.~Sandin, C.~Fuchs, A.~Faessler, H.~Grigorian, G.~R\"opke and J.~Tr\"umper,
  Phys.\ Lett.\ B {\bf 654} (2007) 170.

\bibitem{Benic:2014jia}
  S.~Benic, D.~Blaschke, D.~E.~Alvarez-Castillo, T.~Fischer and S.~Typel,
  Astron.\ Astrophys.\  {\bf 577} (2015) A40.

\bibitem{Kaltenborn:2017hus}
  M.~A.~R.~Kaltenborn, N.~U.~F.~Bastian and D.~B.~Blaschke,
  Phys.\ Rev.\ D {\bf 96} (2017) no.5,  056024.

\bibitem{Blaschke:2013ana}
  D.~Blaschke, D.~E.~Alvarez-Castillo and S.~Benic,
  PoS CPOD {\bf 2013} (2013) 063
  [arXiv:1310.3803 [nucl-th]].

\bibitem{Alvarez-Castillo:2017qki}
  D.~E.~Alvarez-Castillo and D.~B.~Blaschke,
  Phys.\ Rev.\ C {\bf 96} (2017) no.4,  045809.

\bibitem{Bastian:2017fzo}
  N.~U.~F.~Bastian, D.~B.~Blaschke, M.~Cierniak, T.~Fischer, M.~A.~R.~Kaltenborn, M.~Marczenko and S.~Typel,
  arXiv:1710.09189 [astro-ph.HE].
  
\bibitem{Sagert:2008ka}
  I.~Sagert, T.~Fischer, M.~Hempel, G.~Pagliara, J.~Schaffner-Bielich, A.~Mezzacappa, F.-K.~Thielemann and M.~Liebend\"orfer,
  Phys.\ Rev.\ Lett.\  {\bf 102} (2009) 081101
  
\bibitem{Kankiewicz:2017wbo}
  P.~Kankiewicz, M.~Rybczynski, Z.~Wlodarczyk and G.~Wilk,
  arXiv:1710.08193 [hep-ph].
  
\bibitem{VanDevender:2017avk} 
  J.~P.~VanDevender {\it et al.},
  Scientific Reports 7, 8758 (2017).

\bibitem{Klahn:2015mfa}
  T.~Kl\"ahn and T.~Fischer,
  Astrophys.\ J.\  {\bf 810} (2015) no.2,  134.

\bibitem{Gocke:2001ri}
  C.~Gocke, D.~Blaschke, A.~Khalatyan and H.~Grigorian,
  [hep-ph/0104183].

\bibitem{Grigorian:2003vi}
  H.~Grigorian, D.~Blaschke and D.~N.~Aguilera,
  Phys.\ Rev.\ C {\bf 69} (2004) 065802.

\bibitem{Dondi:2016yjl}
  N.~A.~Dondi, A.~Drago and G.~Pagliara,
  EPJ Web Conf.\  {\bf 137} (2017) 09004.

\end{thebibliography}
\end{document}